# Implementing Software Project Control Centers: An Architectural View

Jens Heidrich, Jürgen Münch

Fraunhofer Institute for Experimental Software Engineering,
Fraunhofer-Platz 1, 67663 Kaiserslautern, Germany
{jens.heidrich@iese.fraunhofer.de, juergen.muench@iese.fraunhofer.de}

**Abstract.** Setting up effective and efficient mechanisms for controlling software and system development projects is still challenging in industrial practice. On the one hand, necessary prerequisites such as established development processes, understanding of cause-effect relationships on relevant indicators, and sufficient sustainability of measurement programs are often missing. On the other hand, there are more fundamental methodological deficits related to the controlling process itself and to appropriate tool support. Additional activities that would guarantee the usefulness, completeness, and precision of the resulting controlling data are widely missing. This article presents a conceptual architecture for so-called Software Project Control Centers (SPCC) that addresses these challenges. The architecture includes mechanisms for getting sufficiently precise and complete data and supporting the information needs of different stakeholders. In addition, an implementation of this architecture, the so-called Specula Project Support Environment, is sketched, and results from evaluating this implementation in industrial settings are presented.

## 1    Introduction

Many companies still have problems in setting up effective and efficient mechanisms for project control. According to a study by the Standish Group [1], even though the general expertise in project management and techniques has improved over the last years, around 50% of the projects are still over budget and schedule. Unfortunately, this figure has not changed since the first CHAOS report results were published in 1994. In order to overcome deficits in controlling a software development project, companies have started to introduce so-called software cockpits, also known as Software Project Control Centers (SPCC) [2] or Project Management Offices (PMO) [3], for systematic quality assurance and management support. Software cockpits centrally integrate all relevant information for monitoring and controlling purposes. For instance, a project manager can use them to get an overview of the state of a project, control schedule, effort, and cost, and a quality assurance manager can use them to check the quality of the software produced. An important success factor is that control centers can be customized to the specific goals, organizational characteristics and needs, as well as the concrete project environment. Implementing such control centers is a challenging task. It is not (only) a question of having a customizable generic tool, but primarily a question of finding suitable indicators for controlling the project and having concrete guidelines on how to introduce project controlling functionality and general measurement capabilities into an organization. That is, comprehensive methodological support is needed for successfully setting up and using mechanisms for quantitative project control. There are several approaches for deriving indicators and metrics from high-level measurement goals. One of the most popular ones is the Goal Question Metric (GQM) paradigm [4], which supports explicit definition of measurement goals and has a structured approach for deriving corresponding metrics via a set of questions that help to determine whether the measurement goal has been achieved. However, with respect to project control, a comprehensive methodology that supports the whole life cycle including planning and setting up project control mechanisms, using them continuously for controlling a development project, systematically analyzing the deficits of the used mechanisms, and packaging experiences in order to continuously improve project control, is usually missing.

Specula is a state-of-the-art approach for project control. It interprets and visualizes collected measurement data in a goal-oriented way in order to effectively detect plan deviations. The control functionality provided by Specula depends on the underlying goals with respect to project control. If these goals are explicitly defined, the corresponding functionality is composed out of packaged, freely configurable control components. Specula was mainly developed in the context of the public German research project Soft-Pit (No. 01ISE07A) and makes use of the Quality Improvement Paradigm (QIP) [5] for integrating project control activities into a continuous improvement cycle. Furthermore, the GQM approach is used for explicitly specifying measurement goals for project control. The basic methodology and an extensive usage example are described in [6]. The approach was evaluated as part of industrial case studies in the Soft-Pit project, where the prototypical implementation was used to provide project control functionality for real development projects. Results of the first two iterations can be found in [7] and [8]. A summary of success factors extracted so far from applying the approach and our experience in setting up and using quantitative project control are presented in [9].

The aim of this paper is to talk about how to concretely implement a control center addressing all relevant goals with respect to project control following the general Specula methodology. Section 2 gives an overview of typical problems that have to be addressed when implementing control centers and summarizes strengths and weaknesses of existing methods and technical approaches. Section 3 illustrates a conceptual architecture for control centers and the basic functionality that has to be provided. Moreover, the basics concepts of the Specula approach are summarized, including the conceptual model and the basic methodology for setting up and using the project control functionality. Section 4 presents the Specula Project Support Environment tool, which was implemented based on this architecture and was used as a kind of product line for flexibly composing the needed project control functionality for the different case studies conducted. Section 5 summarizes the results from evaluating the approach, including some lessons learned with respect to the concrete tool prototype used. Section 6 concludes with a brief summary and outlook on future work.

## 2    Project Control in Research and Practice

Setting up a set of suitable mechanisms for project control and applying them correctly during the lifetime of the project is a challenging task. Especially for small and medium-sized enterprises, it is difficult to establish mechanisms for quantitative project control due to the limited resources for setting up appropriate processes and analyzing data. If expert knowledge for setting up a customized measurement program for project control is missing or its implementation seems to be too costly, project control is often done using out-of-the-box functionality as provided by standard project control tools instead of defining and controlling specific measurement goals. Typical dashboards provide only a fixed set of indicators and visualizations with quite simple customization mechanisms; a higher-level quality model that helps to analyze and interpret the indicators in the context of a clearly defined measurement goal is usually missing. There exists a huge set of specific tools for controlling different aspects of cost, time, and quality, but no single point of project control that covers all relevant aspects for controlling the project is provided. In research, several approaches exist that provide partial solutions to the problem of effective and efficient control of development processes. Deficits can be seen especially with respect to supporting purpose- and role-oriented project control by flexibly combining control mechanisms. An overview of these approaches can be found in [2]. The indicators that are used for project control should be derived in a systematic way from the project goals [10] (using, e.g., GQM). Some indicator examples can be found in [11].

In practice, approaches from the business intelligence area, such as Pentaho (*www.pentaho.com/*) or Jaspersoft (*http://www.jaspersoft.com/*) can be used to construct software dashboards. They are able to connect to different data sources, extract the relevant information, and store this information in a database. They offer different analysis engines for providing dashboard visualizations and report generation. They provide an open interface for extending their capabilities towards integrating project control functionality. However, methodological support for systematically deriving the right control mechanisms for a project and organization based on context information and organizational goals is usually missing. Pentaho and Jaspersoft could be customized to address different aspects of project control. Most commercial dashboards in the area of software project control focus on a certain aspect, like technical quality, schedule adherence, or performance indicators. A more holistic approach addressing all aspects relevant for project control is not in the focus of these kinds of dashboards. For instance, the CAST AD Governance Dashboard (*http://www.castsoftware.com/*) focuses on code quality and provides a customizable set of indicators for analyzing and assessing different quality aspects with respect to technical quality. However, it is not clear how to select appropriate indicators that fit the specific goals of a project or a certain organization.

## 3    Conceptual Architecture of Control Centers

Specula (Latin for watchtower) is an approach for constructing control centers in a goal-oriented way. It was developed focusing on extensibility (with respect to the control functionality provided), customizability (with respect to the context in which the control functionality is applied), and reusability (with respect to the functionality offered). It composes the project control functionality out of packaged, freely configurable control components. Specula consists of the following components:

- a *conceptual model* formally describing the interfaces of reusable control components for data collection, data interpretation, and data visualization,
- a *methodology* of how to select control components according to explicitly stated goals and customize the SPCC functionality,
- a *conceptual architecture* for implementing software cockpits, and
- a *prototype implementation* of the conceptual model, including a construction kit of predefined control components.

The conceptual model as well as the basic methodology and a high-level conceptual architecture were presented in [6]. In this section, we will summarize some basics with respect to the model (Section 3.1) and the methodology (Section 3.2) needed to understand the basic structure of the architecture and the corresponding prototype implementation. After that, all elements of the conceptual architecture (Section 3.3) will be discussed. As the focus of this paper is on implementing control centers, the prototype implementation will be discussed in a separate section (Section 4).

### 3.1    Conceptual Model

The central component of the Specula conceptual model is a *visualization catena (VC)*, which defines components for automatically and manually collecting measurement data, processing and interpreting these data, and finally visualizing the processed and interpreted data. The whole visualization catena has to be adapted in accordance with the context characteristics and organizational environment of the software development project currently being controlled. Fig. 1 gives an overview of all VC components and their corresponding types. Specula distinguishes between the following five components on the type level from which a concrete VC is instantiated:

- *Data types* describe the structure of incoming data and data that is further processed by the VC. For instance, a time series (a sequence of time stamp and corresponding value pairs) or a project plan (a hierarchical set of activities having a start and end date and an effort baseline) could be logical data types.
- *Data access object packages* describe the different ways concrete data types may be accessed. A special package may be used, for instance, to automatically connect to an effort tracking system or bug tracking database.
- *Web forms* describe a concrete way of managing measurement data manually, involving user interaction. A web form refers to certain data types that are needed as input. For instance, in order to enter effort data manually, one needs the concrete activities of the project for which the effort is tracked.
- *Functions* represent a packaged control technique or method, which is used to process incoming data (like Earned Value Analysis, Milestone Trend Analysis, or Tolerance Range Checking). A function needs different data types as input and produces data of certain data types as output.
- *Views* represent a certain way of presenting data, like drawing a two-dimensional diagram or just a table with a certain number of rows and columns. A view visualizes different data types and may refer to other views in order to create a hierarchy of views.

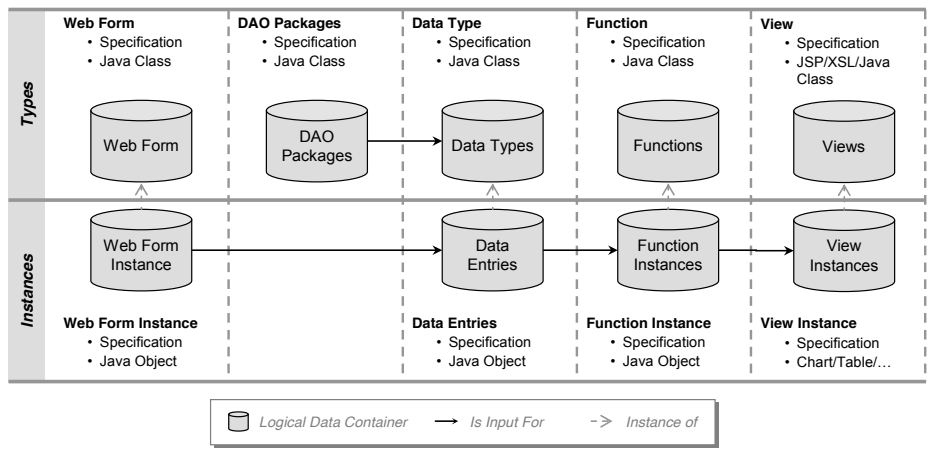

**Fig. 1.** Basic structure of the Specula repository. Each type has a formal specification (e.g., inputs, outputs, parameters) and a corresponding implementation. A type may be instantiated. Such instances also have a formal specification (e.g., the concrete data that is used as input or that is produced, or the concrete parameter setting that is used) and use the implementation of the corresponding type to perform their tasks (e.g., reading, aggregating, or visualizing data).

A VC is instantiated from the types described above by using the following components on the instances level:
- *Data entries* instantiate data types and represent the concrete content of measurement data that are processed by a control center. External data must be read-in or imported from an external location, or manually entered into the system. Each external data object has to be specified explicitly by a data entry containing, for instance, the start and end times and the interval at which the data should be collected. In addition, the data access object package that should be used to access the external data has to be specified.
- *Web form instances* provide web-based forms for manually managing measurement data for data entries.
- *Function instances* apply the instantiated function to a certain set of data entries. A function instance processes data and produces output data, which could be further processed by other function instances or visualized by view instances.
- *View instances* apply the instantiated view to a certain set of data entries. A view instance may refer to other view instances in order to build up a hierarchy.

A visualization catena consists of a set of data entries, each having exactly one active data access object for accessing incoming data, a set of web form instances for managing the defined data entries, a set of function instances for processing data, and finally, a set of view instances for visualizing the processing results.

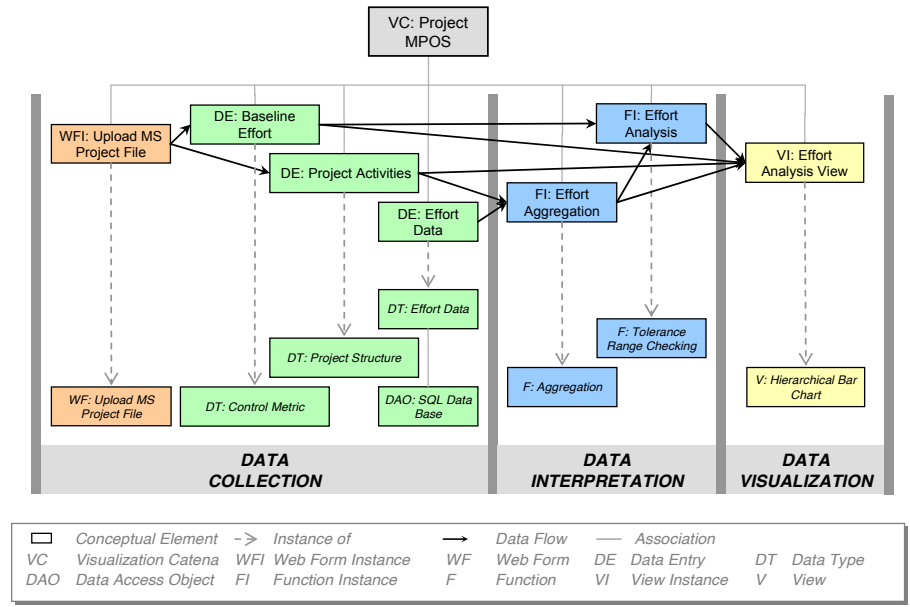

**Fig. 2.** Example of a visualization catena. A visualization catena is composed of web form instances, data entries, function instances, and view instances. These elements are instances of corresponding types: web forms, data types, functions, and views.

Fig. 2 presents excerpts of the visualization catena for a practical course held at the University of Kaiserslautern. The catena contains all control components needed for ensuring that the actual effort of the project stays below the planned effort for all activities. The upper part of the figure shows the instances and the lower part the instantiated types, that is, the reused control components. The data collection area specifies three data entries: one representing the baseline effort per activity (instance of a control metric data type), one representing the hierarchy of project activities (instance of a general project structure data type), and one representing the actual effort data per project team member and project activity (instance of a general effort table and accessed via an SQL data connector). For collecting the project activities and the planned effort, a web form instance is defined, which imports the information from an MS Project file. The data processing area defines two function instances: one for aggregating effort data across the defined activities (in order to compute the actual effort per project activity) and one for comparing the actual effort with the planned effort per activity (making use of a tolerance range checking function). The data visualization area defines one view instance that visualizes the actual effort, the planned effort, and the computed effort deviation along all project activities using a bar chart that is able to drill down into the effort data along the hierarchy of project activities.

### 3.2    Methodology

Specula makes use of the QIP for introducing an improvement-oriented software project control cycle. QIP is used to implement a project control feedback cycle and make use of experiences gathered for reusing and customizing control components. GQM is used to drive the selection process of finding the right control components according to explicitly defined measurement goals. The different phases that have to be considered for setting up and applying project control mechanisms can be characterized as follows (see [6] for a more extensive discussion and examples):

- *I. Characterize Control Environment:* First, project stakeholders characterize the environment in which project control shall be applied in order to set up a corresponding measurement program that is able to satisfy all needs.

- *II. Set Control Goals:* Then, measurement goals for project control are defined and metrics are derived determining what kind of data to collect. In general, any goal derivation process can be used for defining control objectives. For practical reasons, we focus on the GQM paradigm for defining concrete measurement goals.

- *III. Goal-oriented Composition:* Next, all control mechanisms for the project are composed based on the defined goals; that is, control techniques and visualization mechanisms are selected from a corresponding repository and instantiated in the context of the project that has to be controlled. This process is driven by a measurement plan that clearly defines which indicators contribute to specific control objectives, how to assess and aggregate indicator values, and how to visualize control objectives and intermediate results.

- *IV. Execute Project Control Mechanisms:* Once all control mechanisms are specified, a set of role-oriented views is generated for controlling the project. When measurement data are collected, the control mechanisms interpret and visualize them accordingly, so that plan deviations and project risks are detected and a decision-maker can react accordingly. If a deviation is detected, its root cause must be determined and the control mechanisms have to be adapted accordingly. This, does, for example, require data analyses on different levels of abstraction in order to be able to trace causes of plan deviations.

- *V. Analyze Results:* After project completion, the resulting visualization catena has to be analyzed with respect to plan deviations and project risks detected in time, too late, or not detected at all. The causes for plan deviations and risks that were detected too late or that were not detected at all have to be determined.

- *VI. Package Results:* The analysis results of the control mechanisms that were applied may be used as a basis for defining and improving control mechanisms for future projects (e.g., selecting the right control techniques and data visualizations, choosing the right parameters for controlling the project).

### 3.3    Conceptual Architecture

[9] presents a more abstract representation of the conceptual SPCC architecture. The view presented here is more sophisticated and addresses visualization catena handling

in much more detail ([2] presents an earlier version of this view). The SPCC architecture is organized along three different layers. The *information layer* gathers all information and data that are essential for the functionality, for instance measurement data from the current project, experiences from previous projects, and internal information, such as all available Specula instances and types. The *functional layer* performs all data processing activities; that is, it executes chosen function instances and composes view instances. Finally, the *application layer* is responsible for all user interactions; that is, it provides the resulting information of the functional layer to an SPCC user and receives all incoming user requests. Each layer consists of several conceptual elements that provide the essential project control functionality. An overview of the architecture is presented in Fig. 3. In the following, an overview of the essential conceptual elements covered by the conceptual architecture is given.

- *Repository Management Unit:* The repository management unit provides access to a repository containing reusable parts of the underlying conceptual model: VC types (data types, DAO packages, functions, views, and web forms) and VC instances (data entries, function instances, view instances, and web form instances).

- *EB Management Unit:* The experience base management unit provides access to an experience base (EB). One EB section provides project-specific information, such as the measurement data of the current project, the project goals and characteristics, and the project plan. The other EB section provides organization-wide information, such as quality models (e.g., as a basis for data prediction) and qualitative experience (e.g., to guide a project manager by providing countermeasures). The EB management unit organizes access to an experience base by providing mechanisms for accessing distributed data sources (in case of distributed development of software artifacts), for validating incoming data, and for integrating new experiences into the (organization-wide) EB. The EB management unit accesses (external) data sources and creates logical data containers (data entries) that may be used by the data processing and packaging units.

- *Customization Unit:* The customization unit is in charge of creating the visualization catena that is responsible for controlling a software development project. That is, it needs to instantiate the corresponding types from the SPCC repository. The types have to be selected based on the goals and characteristics of the project. Specula uses a GQM plan for specifying measurement goals, questions, and metrics. Based on the information provided there, suitable types are selected from the repository and instantiated. If the SPCC repository does not provide appropriate components, new types have to be defined and stored in the repository that may be reused by future projects. The VC instances have to be customized according to the project specifics. This includes setting the required input and all parameters needed for using the specific type.

- *Data Processing Unit:* The data processing unit receives the visualization catena from the customization unit. It analyzes all function instances, that is, it determines input and output information, the function's implementation, and the relationships to other function instances. If a function instance is based on other function instances, an appropriate execution sequence is computed. During execution of the chosen function instances, the data processing unit receives data entries from the EB management unit, respectively already processed data from a previously exe-cuted function instance. The results of a function instance have to be updated if an underlying data unit or function instance result has changed. The results of all executed functions are delivered to the presentation unit for data visualization.

- *Presentation Unit:* The presentation unit receives the visualization catena from the customization unit. It analyzes all view instances, that is, it determines the relationships between the view instances and the function instance outputs and data entries that have to be used to create the corresponding visualization. If a view instance is based on other view instances, an appropriate creation sequence is computed. A view instance has to be updated if the underlying data has changed. The results of all views are delivered to the user communication unit.

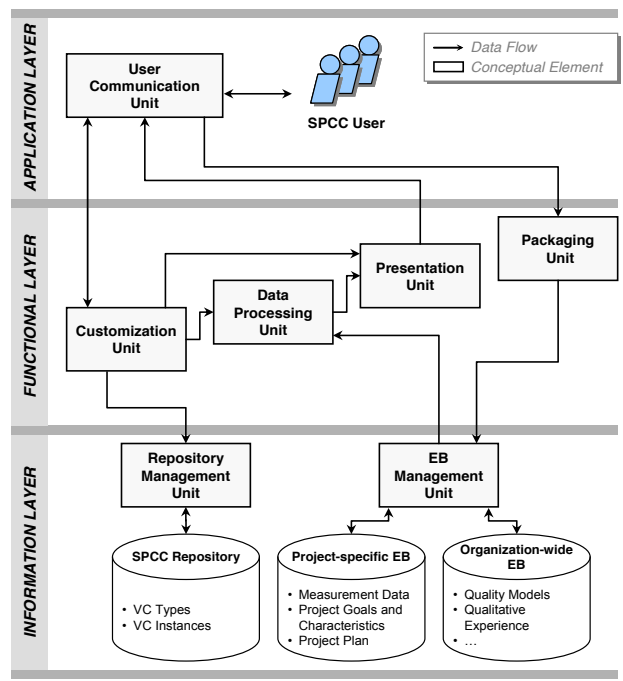

**Fig. 3.** General conceptual architecture of the Specula project control center. The different conceptual elements represent logical tasks that need to be performed by a control center in order to access information, interpret and analyze it, and communicate with an SPCC user.

- *Packaging Unit:* The packaging unit is responsible for all information that is fed back into the system by the user. This includes all external data provided via web form instances. It summarizes all experiences gained from the usage of an SPCC, adapts them according to the needs of future projects (i.e., generalizes the information units), and delivers them to the EB management unit for integration into the respective section of an experience base.
- *User Communication Unit:* The user communication unit determines the access granted to a specific user. That is, it permits a certain user to access the results of a certain set of function instances or a certain set of view instances. Furthermore, it provides a graphical user interface (GUI) for administering the SPCC (e.g., user management) and for the goal-oriented selection of VC components (via the customization unit). This includes selecting appropriate data types, functions, views, and web forms, and adapting the resulting visualization catena. Last, it provides access to the generated visualizations (delivered by the presentation unit) and manages interaction with them (e.g., drilling down or filtering data).

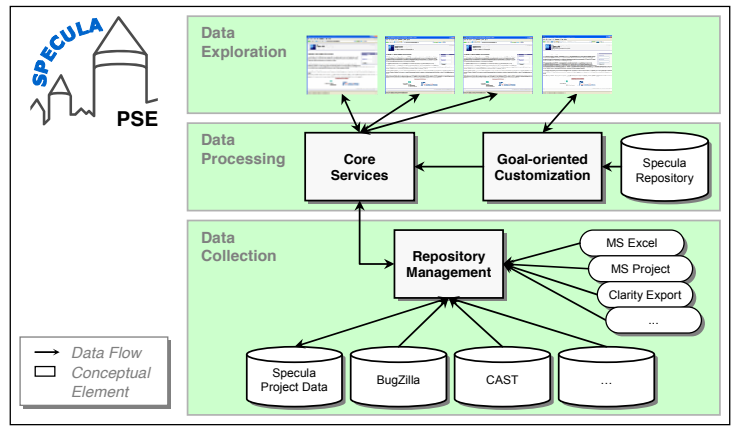

**Fig. 4.** Technical high-level architecture of the Specula PSE tool. The tool provides capabilities for collecting data, processing them according to the specified visualization catena, and finally visualizing and exploring them.

## 4   The Specula Project Support Environment Tool

Large parts of the conceptual architecture presented above are implemented by the Specula Project Support Environment (PSE) tool, which completely automates the conceptual units except for parts of the customization and packaging units. The Specula PSE tool can be used as a framework for systematically composing project control mechanisms based on reusable control components; it provides a core functionality for project control and clearly defines interfaces for specifying additional modules that can be freely enhanced with respect to specific needs. Customization includes specification of types, instances, and administration information (users and groups), implementation of data access object packages for accessing different repositories, implementation of data types for defining logical data containers, implementation of functions for processing measurement data, implementation of views for displaying data, and implementation of web forms for managing (importing, exporting, adding, removing) data. Specula PSE is a web-based software implemented as a Java-Servlet and runs on top of a Tomcat web server. The tool has a classical three-layered design (as presented in Fig. 4) in correspondence to the layers of the conceptual architecture:

- The *data collection layer* deals with accessing different data sources. Project data and measurement data need to be collected automatically by accessing different ex-isting databases, or semi-automatically by using web forms for importing data from files or for entering data manually. For instance, a data type and corresponding data access object may be specified for accessing defect data stored in a BugZilla database (*http://www.bugzilla.org/*), or a web form may be specified for importing project plan information stored in an MS Project file.

- The *data processing layer* uses the data collection layer for accessing data from different sources in a unique way, processing them according to the VC defined, and finally providing services upon the processing results. Different services are offered for user management, checking the consistency of a VC specification, accessing data repositories and VC specifications, etc. In order to adapt the Specula PSE functionality to project goals and characteristics, a corresponding customization unit manages all control components; that is, it supports the definition of new control components, the reuse of existing components, and the parameterization of control components according to the project context.

- The *data exploration layer* uses the services of the data processing layer for providing a graphical user interface, including displaying charts and tables, managing data, administering control components, and importing/exporting data.

The process of deriving a VC from a GQM plan (including project goals and characteristics) is currently not automated by the tool and must be performed manually. In the future, this process could partly be automated depending on the degree of formality of the corresponding GQM models, interpretation models, and further contextual information. However, currently, performing this process requires a deeper understanding of the measurement program and the control components of the Specula repository that may potentially be reused for implementing the measurement program. The Specula prototype tool automates the specification and packaging of all control components of the conceptual model and is able to automatically execute the derived visualization catena. SPCC users may use the tool for collecting measurement data and for utilizing the generated visualizations for project control. Support for setting up and accessing an organizational experience base is currently also limited and restricted to managing control components. The control components contained in the Specula repository depend on the organization (and the very project that should be controlled). Some components may be more general and applicable for several companies and projects, whereas others may be very specific and implement organization-specific control strategies. This is also related to the different *kinds* of components in the repository. For instance, one control component may implement a (fairly) complex control technique (like Earned Value Analysis) and another component may just provide some simple data processing functionality for supporting other functions (like scaling a time series or converting between different data types). Specula PSE comes with a set of standard data collection forms, control techniques, and views that were used as part of case studies and may serve as a basis for adding further elements to the framework.

Fig. 5 shows the internal structure of the tool. Let us assume that the VC as shown in Fig. 2 was specified and executed by the tool. Let us also assume that the project plan was updated and a user wants to import an MS Project file by using a web form instance of the VC. Web form instances are implemented as Java Server Pages (JSP). Based on the VC specification, the tool automatically creates a web page for upload-ing the file. During uploading, the content of the file is analyzed and then transformed into so-called transport objects (implemented as Java classes). The three layers of the system are connected via these transport objects. They actually contain the data that is collected, processed, and visualized. The content of the transport objects is stored in the system using the repository service. According to the VC, the file contains (a) a list of project activities and (b) the baseline effort for all activities. The VC manager recognizes that data belonging to two data entries of the VC was updated and automatically initiates an update of the corresponding function instances (and other control components affected). In our case, the implementation of the function instance, comparing the actual effort against the baseline effort, is invoked and the effort analysis is performed according to its specification. That is, a corresponding Java class is instantiated, provided with the necessary data for performing the effort analysis, and invoked accordingly.

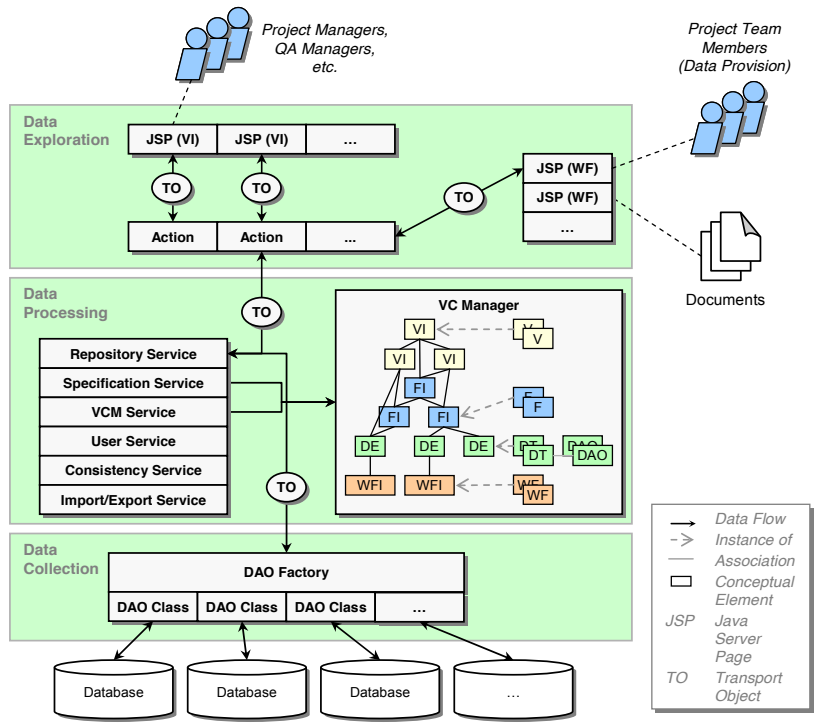

**Fig. 5.** Internal structures and data flow of the Specula PSE tool. The data collection layer accesses different data repositories and sends the information via so-called transport objects to the data processing layer and the VC manager, which in turn processes the data according to the VC specification and passes it on to the data exploration layer for visualization. Transport objects may also flow back to the back end layer if data is imported from other sources.

View instances can be implemented in three different ways. The most common way is to provide a JSP page containing some graphical illustrations (using, e.g., JFreeChart, *http://www.jfree.org/*) and tables. Based on the VC specification, the tool automatically creates a web page for displaying the graphical representation of the view. If the project manager wants to see the effort controlling view of the example VC, the corresponding JSP page is provided with the necessary data for creating a bar chart containing the planned and actual effort per project activity. Buttons for navigating through the hierarchy of project activities and filtering data are also provided. The required data is automatically retrieved from a project database, stored in a transport object, and delivered to the exploration layer using the repository service.

## 5     Evaluation Results

The approach was evaluated as part of industrial case studies in the Soft-Pit project in which the prototypical implementation was used. Results of the first two iterations can be found in [7] and [8]. In this section, we will highlight some of the evaluation results over the three project iterations. In general, people perceived the usefulness and ease of use of the Specula control center as positive (which was evaluated using the Technology Acceptance Model [12]). The results for ease of use were not as promising as the results for usefulness. This is not a surprising result given the fact that a prototype was used during the case studies. All users received a basic training in using the control center, but depending on their familiarity with such tools, the results varied. The general usefulness and ease of use also varied across the different case study providers depending on the state of the practice before introducing the Soft-Pit control center solution. We continuously improved the method for setting up the control center and provided concrete guidelines for the case study providers on how to perform concrete tasks. As a consequence, the usefulness increased continuously over the three iterations. For evaluating the efficiency of the control center, we analyzed the detected plan deviations and project risks. Overall, 18 deviations and risks were detected by the control center in the second and 21 in the third iteration. The approach was able to detect between 40% and 80% of the listed plan deviations and project risks earlier than the traditional approaches to project control used by the case study providers before introducing the Soft-Pit solution. More than 20% of plan deviations and project risks were found that would not have been detected at all without using the control center. The contexts in which the control center was applied differed quite a lot depending on the case study provider. It included small and medium-size companies as well as a large organization. The ratio of control center costs to the overall development costs varied between 11% and 14% for a team size of 7 team members and between 9% and 10% for a team size of 17 team members. This relatively high ratio might have been related to the fact that some tasks had to be performed manually, and that the evaluation period was too short, so that activities that usually have to be performed just once had a bigger impact on the overall figures.

## 6     Conclusions

This article presented a conceptual architecture for control centers and the Specula PSE controlling tool implementing this architecture. The approach implements a dynamic approach for project control; that is, measures and indicators are not prede-termined and fixed for all projects. They are dynamically derived from measurement goals at the beginning of a development project. A context-specific construction kit is provided, so that elements with a matching interface may be combined. The qualitative benefits of the approach include: allowing for more transparent decision-making, reducing the overhead for data collection, increasing data quality, and, finally, achieving projects that are easier to plan and to control. Future work will concentrate on setting up a holistic control center that integrates more aspects of engineering-style software development. The starting point for setting up such a control center are usually high-level business goals, from which measurement programs and controlling instruments can be derived systematically. Thus, it would be possible to transparently monitor, assess, and optimize the effects of business strategies.

## References

1. Standish Group. CHAOS Summary 2008. Study, Standish Group International, 2008.
2. Münch, J., Heidrich, J.: Software Project Control Centers: Concepts and Approaches. *Journal of Systems and Software* **2004; 70 (1)**, pp. 3-19.
3. Project Management Institute: *A Guide to the Project Management Body of Knowledge (PMBOK® Guide)* 2000 Edition. Project Management Institute, Four Campus Boulevard, Newtown Square, PA 19073-3299 USA, 2000.
4. Basili, V.R., Caldiera, G., Rombach, D: Goal Question Metric Approach. *Encyclopedia of Software Engineering*, John Wiley & Sons, Inc., 1994, pp. 528-532.
5. Basili, V.R., Caldiera, G., Rombach, D: The Experience Factory. *Encyclopaedia of Software Engineering 1*, John Wiley & Sons, Inc., 1994, pp. 469-476.
6. Heidrich, J.; Münch, J.: Goal-oriented setup and usage of custom-tailored software cockpits. In: *Proceedings of the 9th International Conference on Product-focused software process improvement (PROFES 2008)*, Monte Porzio Catone, Italy, June 23-25, 2008, pp. 4-18.
7. Ciolkowski, M., Heidrich, J., Münch, J., Simon, F., Radicke, M.: Evaluating Software Project Control Centers in Industrial Environments. *International Symposium on Empirical Software Engineering and Measurement (ESEM 2007)*, Madrid, 2007, pp. 314-323.
8. Ciolkowski, M., Heidrich, J., Simon, F., Radicke, M.: Empirical Results from Using Custom-Made Software Project Control Centers in Industrial Environments. *International Symposium on Empirical Software Engineering and Measurement (ESEM 2008)*, Kaiserslautern, 2008 (to be published).
9. Ciolkowski, M.; Heidrich, J.; Münch, J.: Practical guidelines for introducing software cockpits in industry. In: *Proceedings of the 5th Software Measurement European Forum, (Smef 2008)*, Milan, May 28-29-30 2008, pp. 49-64.
10. Kitchenham, B.A.: *Software Metrics*. Blackwell, Oxford, 1995.
11. Agresti, W., Card, D., Church, V.: *Manager's Handbook for Software Development*. SEL 84-101, NASA Goddard Space Flight Center. Greenbelt, Maryland, November 1990.
12. Davis, F.D.: Perceived usefulness, perceived ease of use, and user acceptance of information technology. *MIS Quarterly* **1990; 13(3)**, pp. 319-340.